\documentclass[aps,prd,preprint]{revtex4}

\usepackage{dcolumn}
\usepackage{amsmath}

\usepackage{graphicx}
\usepackage{epstopdf}
\def\a{\alpha}
\def\b{\beta}

\def\e{\epsilon}                

\def\p{\pi}                     
\def\s{\sigma}                  

\def\D{\Delta}

\def\co{{\cal O}}

\long \def \blockcomment #1\endcomment{}

\newcommand{\bee}{\begin{equation}}
\newcommand{\ee}{\end{equation}}
\newcommand{\beea}{\begin{eqnarray}}
\newcommand{\eea}{\end{eqnarray}}

\begin{document}
\title{Investigating the  critical properties of beyond-QCD theories using
Monte Carlo Renormalization Group matching}

\author{Anna Hasenfratz}
\email{anna@eotvos.colorado.edu}
\affiliation{Department of Physics,
University of Colorado, Boulder, CO 80309, USA}

\begin{abstract} 
Monte Carlo Renormalization Group (MCRG) methods were designed  to study the non-perturbative phase structure and critical behavior of
 statistical systems and quantum field theories. I adopt the 2-lattice matching method used extensively in the 1980's 
 and show how it can be used to predict the 
 existence of non-perturbative fixed points and their related critical exponents in many flavor SU(3) gauge theories. This work serves to test the method and I study relatively well understood systems: the $N_f=0$, 4 and 16 flavor models. The  pure gauge and  $N_f=4$ systems are confining and chirally broken and the MCRG  method can predict their bare step scaling functions. Results for the $N_f=16$  model  indicate the existence of an infrared fixed point with nearly marginal gauge coupling. 
 I present preliminary results for the scaling dimension  of the mass at this new fixed point.

\end{abstract}
\maketitle

\section{Introduction}

The Monte Carlo Renormalization Group (MCRG) methods, based on 
Wilson's renormalization group theory, were developed and used extensively
in the 1980's to study the critical properties of spin and gauge models
\cite{Swendsen:1979gn,Swendsen:1981rb,Swendsen:1984vu,Hasenfratz:1984bx,Hasenfratz:1984hx,Bowler:1984hv,Hasenfratz:1984ju,Hasenfratz:1987zi,Decker:1987mu,Gupta:1984gq,Patel:1984dw}.
The 2-lattice matching MCRG proved to be particularly useful to calculate
the $\beta$ function of asymptotically free theories, like quenched
QCD \cite{Bowler:1984hv,Hasenfratz:1984bx,Hasenfratz:1984hx}. The approach has all
but been forgotten in the last 20 years as lattice QCD calculations
focused on spectral and other experimentally measurable quantities.
Lately there has been increased interest in beyond-QCD lattice models
\cite{Catterall:2007yx,Appelquist:2007hu,Shamir:2008pb,Deuzeman:2008sc,DelDebbio:2008zf,
Catterall:2008qk,Fodor:2008hm,DelDebbio:2008tv,DeGrand:2008kx,Hietanen:2008mr,
Fodor:2008hn,Appelquist:2009ty,Hietanen:2009az,Deuzeman:2009mh}
as they could describe strongly coupled beyond-Standard Model physics 
\cite{Hill:2002ap,Banks:1981nn,Sannino:2004qp,Dietrich:2006cm}. 
Ref. \cite{Fleming:2008gy}  is a good summary of the issues and
recent lattice results. 
A basic discussion of the physical picture  with some remarks on expectations for lattice simulations
were presented in Ref. \cite{DeGrand:2009mt}.  In this paper I follow the Wilson renormalization
group RG language used in \cite{DeGrand:2009mt}. 
 
 SU(3) gauge models with $N_f$ fermions in the fundamental representation can have very different phase structure depending on the number of fermions \cite{Banks:1981nn}. If $N_f>16$ asymptotic freedom is lost, the gauge coupling is irrelevant and the continuum theory is free. For small $N_f$ the $g=0$, $m=0$ Gaussian fixed point (GFP) is the only critical fixed point (FP). The theory is asymptotically free, confining and chirally broken. Somewhere around $N_f\approx10$ the gauge coupling develops a new FP at $g^*\ne0$ \cite{Caswell:1974gg,Dietrich:2006cm}. At $g^*$ the gauge coupling is irrelevant, it is an infrared FP (IRFP). The continuum limit defined in the basin of attraction of this IRFP is neither confining not chirally broken; it is conformal when $m=0$. This conformal phase is expected to exits all the way to $N_f=16$. Identifying the lower end of the conformal window and the critical properties of the IRFP are the main issues of recent lattice simulations.
The MCRG
method was designed to answer these kind of questions and in this
paper I present the first such study in $N_{f}=4$ and $N_{f}=16$
flavor SU(3) theories. I also investigate the pure gauge SU(3) model
where it is possible to do high statistics, large volume simulations.
I have chosen these models as the expected phase structure
is rather well known, so I can use them to calibrate and test the method. 
My eventual goal is to extend these studies to other flavor numbers or fermions in
different representations.

Since MCRG has been used very little in the last 20 years, I devote Sect.\ref{sect:MCRG} to 
the basic description of the 2-lattice matching MCRG. The method allows the determination 
of  a sequence of couplings  $\beta_0,\beta_1,...\beta_n,...$ with lattice spacings 
that differ by a factor of s between consecutive points,
$a(\beta_n)=a(\beta_{n-1})/s$.   $s$ is the scale change of the RG transformation, $s=2$ in this study.
This sequence is  analogous to the step scaling function defined in the Schrodinger functional (SF) method 
\cite{Luscher:1992zx,Luscher:1993gh,Capitani:1998mq},
 but in MCRG it is defined through the bare couplings. 
 To emphasize this difference I will use the notation
$s_b(\beta_n;s)=\beta_n-\beta_{n-1}$ for the bare step scaling function instead of the more traditional $\s(u;s)$ used in the SF approach. 

The 
 sequence $\beta_0,\beta_1,...\beta_n,...$ can be used to determine the renormalized 
running coupling in theories that are governed by the GFP 
if at the weak coupling end of the chain a renormalized coupling, like the SF $\bar{g}^2$,
is calculated and connected to a continuum regularization scheme, while at the strong coupling end some physical quantity is used to
determine the lattice scale. 
I do not pursue this calculation here, though I will compare results for $s_b(\b;s=2)$ from SF and MCRG in Sect.
\ref{sect:pure_SU3}.

The rest of the paper is organized as follows. Sect.  \ref{sect:PT} summarizes the perturbative picture of these many fermion theories. 
Sect. \ref{sect:MCRG} describes the 2-lattice matching method and defines the RG block transformation used in this work. The numerical simulations and results are discussed in Sect. \ref{sect:simulations}. The technical aspects of MCRG are described in  detail for the pure gauge SU(3) theory in Sect.\ref{sect:pure_SU3} as it serves to justify the approach used for the $N_f=4$ and $N_f=16$ models in Sects. \ref{sect:nf4} and \ref{sect:nf16}.  I use nHYP smeared staggered fermions in this work. I present some basic properties of 4 flavor staggered fermions, together with the MCRG calculation of the bare step scaling function in Sect. \ref{sect:nf4}.  The existence of an IRFP requires the re-evaluation of the MCRG method. This, together with  preliminary results for the scaling exponent of the mass in the $N_f=16$ model are presented in Sect. \ref{sect:nf16}.


\section{The perturbative picture \label{sect:PT}}

Before discussing the MCRG method and the numerical results  I  briefly summarize the perturbative picture.
The universal 2-loop $\beta$ function for SU(3) gauge with $N_f$ fermions in the fundamental representation is 
\bee
\beta(g^{2})=\frac{dg^{2}}{d\log(\mu^{2})}=
\frac{b_{1}}{16\pi^{2}}g^{4}+\frac{b_{2}}{(16\pi^{2})^{2}}g^{6}+\dots\,,
\label{eq:betaQCD}
\ee
\beea
b_{1} & = & -11 + \frac{2}{3} N_f\,, \nonumber \\
b_{2} & = & -102+\frac{38}{3} N_f\,. \nonumber 
\eea

For $N_f<16.5$ the 1-loop coefficient $b_1$ is negative, the gauge coupling is relevant at the $g=0$ Gaussian FP,
  the theory is asymptotically free. Dimensional transmutation is responsible for mass generation. The energy  scale
  changes by a factor of 2 between couplings $g_1$ and $g_2$ if
  \bee
  \rm{ln}(2) = -\,\int_{g_1}^{g_2} \frac{g}{\b(g^2)} dg  \,.
  \label{eq:sb_int}
  \ee
  At one loop level this leads to a constant shift in $\b=6/g^2$ and the bare step scaling function is
  \bee
  s_b(\b_1;s=2) = \b_1-\b_2 = -\,\frac{3 \,\rm{ln}(2)}{4\p^2} b_1\,\,\,\,\,\,
(\rm{1-loop}).
 \label{eq:db_pert}
  \ee
 
  For small fermion numbers 
the higher order terms are small, the $\beta$ function is expected to remain negative. Lattice simulations indicate that for
$N_f\le8$ the system is confining and chiral symmetry is spontaneously broken. For $N_f>8$ the 2-loop $\beta$ function
develops a zero at $g^*\ne0$ Banks-Zaks FP\cite{Banks:1981nn}. 
At this new FP $g$ is irrelevant, it is an IRFP for the gauge coupling. The infinite cut-off limit in the vicinity
of $g^*$ is conformal. 

When the perturbatively predicted $g^*$ is large, higher order or non-perturbative effects can destroy the existence of the IRFP. Analytical considerations and numerical simulations suggest  that the bottom of the conformal window is around $N_f\approx10$
\cite{Caswell:1974gg,Dietrich:2006cm}.  At $N_f=16$, the largest flavor number that is
still asymptotically free, the Banks-Zaks FP occours at a small value $g^*\sim \e$, perturbation theory could correctly describe the conformal phase. 
At the IRFP there is only one relevant operator, the mass. Its scaling dimension (critical exponent)  is close to its engineering one  $y_m\sim 1 + \cal{O}(\e)$, while the  scaling dimension of the gauge coupling is $y_g\sim -\e^2$. The slope of the $\b$ function at $g^*$ predicts the exponent
\beea
\b(g^2) & = &-y_g (g^2-g^{*2}) + \co ((g^2-g^{*2})^2) \,,\\
y_g & = &-\frac{b_1^2}{b_2}\,.
\label{eq:gamma_g}
\eea
Eq. \ref{eq:sb_int} now gives
\beea
g_1^2-g^{*2} &=& (g_2^2-g^{*2}) 2^{-2y_g}\,,\\
s_b(\b_1;s=2) &=& \b_1-\b_2 = (\b_2-\b^*)(2^{-2y_g}-1)\,,
\label{eq:db_IRFP}
\eea
if $\b_1-\b^*\,,\,\b_2-\b^* \ll 1$. 
For 16 flavors perturbatively $y_g\approx -0.01$, the gauge coupling is almost marginal.  For smaller $N_f$ or higher representation fermions  
$|y_g|$ can be larger, though both numerical and analytical considerations find that $|y_g|$  remains small even at the bottom of the conformal window\cite{Gardi:1998ch,Appelquist:2009ty,DeGrand:2009mt}.

The mass is a relevant operator both at the GFP and at the IRFP, with critical value $m^*=0$. Under a scale change $s=2$ it changes as
\bee
m_1=m_2 2^{-y_m}\,,
\label{eq:mass_scale}
\ee
where $1/y_m=\nu$ is the critical index of the mass.


\section{The MCRG method  \label{sect:MCRG}}

The Wilson RG description of statistical systems is a very effective
approach to describe the phase diagram, calculate critical indices,
and in case of lattice discretized quantum field theories, understand
the infinite cut-off continuum limit of these models.  
There are many books and review articles written about the subject. I do not attempt 
to explain Wilson RG here, I only  summarize the main points. Two reviews that could be useful for 
other parts of this paper are Refs. \cite{Hasenfratz:1984ju,DeGrand:2009mt}. 

In the inherently
non-perturbative Wilson RG approach one considers the evolution
of all the possible couplings  under an RG transformation
that preserves the internal symmetries of the system but integrates
out the cut-off level UV modes. The fixed points of the transformation
are characterized by the number of relevant  operators,
i.e. couplings with positive scaling dimensions that flow away from the FP. 
Irrelevant couplings have negative scaling dimensions and they flow towards the
FP. The  IR  values of irrelevant operators 
are independent of their UV values. Continuum (or infinite
cut-off) limits can be defined by tuning  the relevant couplings 
towards the FP, thus controlling their IR value. The number of relevant operators
and their speed along the RG flow lines are universal,
related to the infrared  properties of the underlying continuum limit.
On the other hand the location of the FP is not physical, in fact different
RG transformations have different fixed points. 

In quantum field theories the best understood fixed points are at vanishing couplings
(Gaussian FPs) as they can be treated perturbatively. For example the GFP of the 4
dimensional SU(3) pure gauge model has one relevant operator, the
gauge coupling, and no other FP of the model is known to exist. The
Gaussian FP of 2-flavor QCD has two relevant operators, the mass and
the gauge coupling. Gauge theories with many flavors can develop, in addition to the  GFP, a
new fixed point where only the mass is relevant (Banks-Zaks  infrared
fixed point) \cite{Banks:1981nn}. These new FPs are rarely in the perturbative region and to study their existence 
and properties  is the main motivation for this paper.

\subsection{The 2-lattice matching MCRG method}

Consider a $d$-dimensional lattice model with action $S(K_{i})$.
$\{K_{i}\}$ denotes the set of all possible couplings, though in
a typical lattice simulation only a few of them are non-zero. The
system is characterized by one or more length scales, like the correlation
length $\xi$, inverse quark masses, etc. In numerical simulations we always
deal with finite volume and for now I assume a hypercubic geometry with
linear size $L=\hat{L}a$. The first step of a real space renormalization
group block transformation is to define block variables.
These new variables are defined as some kind of  local average of
the original lattice variables and for a scale $s>1$ transformation
they live on an $\hat{L}/s$ lattice. By integrating out the original
variables while keeping the block variables fixed one removes the
ultraviolet fluctuations below the length scale $sa$. The action
that describes the dynamics of the block variables is usually much
more complicated than the original one, but if $s$ is much smaller
than the lattice correlation length, the long distance infrared properties
of the system are unchanged. After repeated block transformation steps
the blocked actions describe a flow line in the multi-dimensional action
space
\bee
\{K_{i}\}\equiv\{K_{i}^{(0)}\}\to\{K_{i}^{(1)}\}\to\{K_{i}^{(2)}\}\to....\,,
\ee
where $\{K_{i}^{(n)}\}$ denotes the couplings after $n$ blocking
steps. While the physical correlation length is unchanged, the lattice correlation 
length after $n$ blocking steps is
\beea
\xi^{(n)} & = & s^{-n}\xi^{(0)}\,.
\label{eq:corr_length}
\eea
 The RG can have fixed points only when $\xi=\infty$ (critical) or
$\xi=0$ (trivial). We are, of course, interested in the former one.
Near the critical fixed point the linearized RG transformation predicts
the scaling operators and their corresponding scaling dimensions.

\begin{figure}
\includegraphics[width=0.65\textwidth,clip]{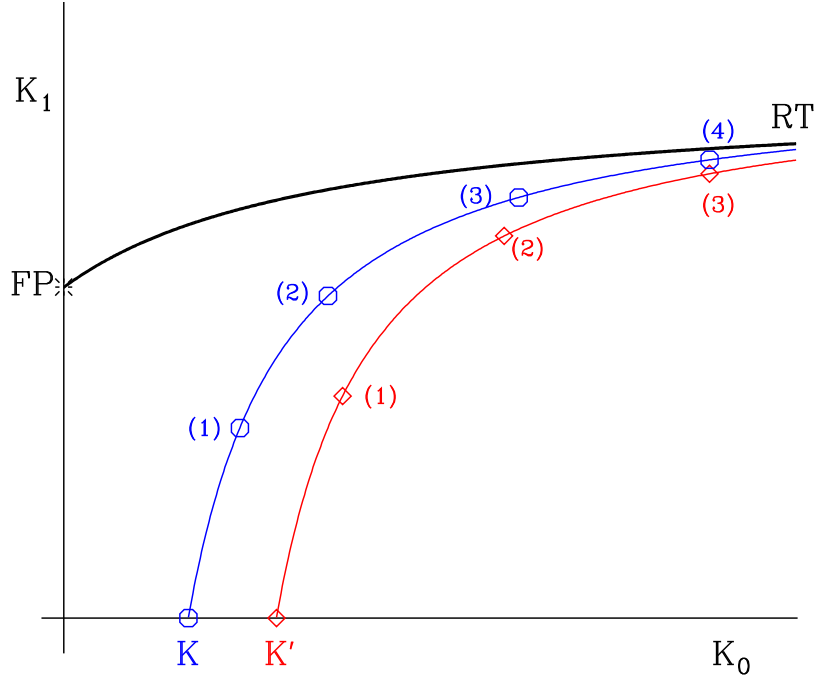}
\caption{Sketch of the RG flow around a FP with one relevant operator. The coupling pair $(K,K')$ 
indicates matched couplings whose correlation length differ by a factor of $s$. \label{fig:RG-flow}}
\end{figure}

It is easy to visualize the renormalization group flow lines when
there is only one relevant coupling at the fixed point, as illustrated
in Fig. \ref{fig:RG-flow}. The sketch depicts the flow lines in the
parameter space $\{K_{0},K_{1}\}$, where for simplicity I assume
that the critical surface is at $K_{0}=0$. Flow lines starting near
the critical surface approach the fixed point in the irrelevant direction(s)
but flow away in the relevant one. After a few RG steps the irrelevant
operators die out and the flow follows the unique renormalized trajectory
(RT), independent of the original couplings. If we can identify
two sets of couplings, $\{K_{i}\}$ and $\{K_{i}^{\prime}\}$, that
end up at the same point along the RT after repeated blocking steps,
we can conclude that their correlation lengths are   identical. If they end
up at the same point along the RT but one requires one less blocking
steps to do so, according to Eq. \ref{eq:corr_length}   their lattice correlation lengths
differ by a factor of $s$. This is also illustrated in Fig. \ref{fig:RG-flow}.
From $K'$ one needs 3 while from $K$ one needs 4 RG steps to reach
the same point of the RT (up to small corrections in the irrelevant
direction(s)), therefore $\xi'=\xi/s$. This gives the bare step scaling
function with scale change $s$. The two lattice matching \cite{Hasenfratz:1984hx,Hasenfratz:1984bx} is
a numerical method to identify $(K,K')$ pairs.

In order to identify a pair of couplings $(K,K^{\prime})$ with $\xi'=\xi/s$
we have to show that after $n$ and $(n-1)$ blocking steps their
actions are identical, $S(K_{i}^{(n)})=S(K_{i}^{\prime(n-1)})$. It
is quite difficult to calculate the blocked action, but fortunately
we do not need to know the actions explicitly to shows that they are
identical. It is sufficient to show that the expectation values of
every operator measured on configurations generated with one or the
other action are identical. Furthermore it is possible to create a configuration
ensemble with Boltzman weight of an RG blocked action by generating
an ensemble with the original action and blocking the configurations
themselves \cite{Swendsen:1979gn}. This suggests the following procedure for the 2-lattice
matching:

\begin{enumerate}
\item Generate a configuration ensemble of size $L^{d}$ with action $S(K)$.
Block each configuration $n$ times and measure a set of expectation
values on the resulting $(L/s^{n})^{d}$ set.
\item Generate configurations of size $(L/s)^{d}$ with action $S(K')$,
where $K'$ is  a trial coupling. Block each configuration $n-1$
times and measure the same expectation values on the resulting $(L/s^{n})^{d}$
set. Compare the results with that obtained in step 1. and tune the
coupling $K'$ such that the expectation values agree.
\end{enumerate}
A few basic comments are in order:

\begin{itemize}
\item [{a)}] Since we always compare measurements on the same lattice size, the
finite volume corrections are minimal and even very small lattices
can be used. 
\item[{b)}] It is not necessary to work
on lattices that are larger than the correlation length of the system,
nor does it matter if we are in the confined or deconfined phase of
the system. 
\item [{c)}] If the flow lines follow the unique RT, even one operator
expectation value is sufficient to find the matching coupling, all
other operators should give the same prediction. In practice we can
do only a few blocking steps and the flow lines might not reach the
RT. That will be reflected by different operators predicting different
matching values. The spread of these predictions measure
the goodness of the matching. Increasing the number of blocking steps
improves the matching, and when the RT is reached, consequetive blocking
steps predict the same matching couplings.
\item [{d)}] The location of the fixed point and its renormalized trajectory
in the irrelevant directions depend on the block transformation. Block
transformations that have free parameters can be optimized so their
RT is reached fast and the matching is reliable after a very few RG steps.
This optimization proved essential in previous applications \cite{Hasenfratz:1984bx,Hasenfratz:1984hx,Bowler:1984hv,Hasenfratz:1987zi,DeGrand:1995ji}.
\item [{e)}] Since we can match by comparing local operators, the statistical
accuracy is usually acceptable even with small configuration sets. 
\end{itemize}

If the FP has two relevant operators, the matching proceeds similarly
but one has to tune 2 operators. In practice this is much more difficult
than the tuning of a simple coupling. It is frequently  easier to fix one of the relevant couplings to its FP value and proceed with the matching in the second relevant coupling as described above.

I will illustrate the above points in Sect.\ref{sect:simulations}.

\subsection{The renormalization group block transformation}

I chose a scale $s=2$ block transformation, similar to what was used
in Refs. \cite{Swendsen:1981rb,Hasenfratz:1984hx} 
\begin{equation}
V_{n,\mu}={\rm Proj[}(1-\alpha)U_{n,\mu}U_{n+\mu,\mu}+\frac{\alpha}{6}\sum_{\nu\ne\mu}U_{n,\nu}U_{n+\nu,\mu}U_{n+\mu+\nu,\mu}U_{n+2\mu,\nu}^{\dagger}]\,,\label{eq:block-trans}
\end{equation}
where ${\rm Proj}$ indicates projection to $SU(3)$. The parameter
$\alpha$ is arbitrary and can be used to optimize the blocking. The
block transformation  used in Refs. \cite{Hasenfratz:1984hx,Bowler:1984hv,DeGrand:1995ji} had
$\a$ fixed, $1-\alpha=\alpha/6$, but instead of projecting to SU(3) the blocked
link was allowed to fluctuate around $V_{n,\mu}$, depending on a free parameter.
In my experience the two block transformations are very similar.

In principle one can define an RG transformation for fermions as well.
However it is easier to do the RG transformation after the fermions
are integrated out, i.e. when the action depends on the gauge fields only.

The role of the parameter $\alpha$ is to optimize the block transformation.
While the critical surface of a system is well defined, the location
of the fixed point itself is not physical, it can be changed by changing
the RG transformation. It is important to optimize the blocking so
its FP and RT  can be reached in a few steps. The optimal blocking is characterized
by 

\begin{enumerate}
\item Consistent matching between the different operators: along the RT  all
expectation values should agree on the matched configuration sets. Any
deviation is a measure that the RT has not been reached.
\item Consecutive blocking steps should give the same matching coupling.
When they predict different values, one can try to extrapolate to
the FP using the first non-leading critical exponent. 
\end{enumerate}

In the next Section I will show that both of the above  conditions can be 
satisfied in numerical simulations if the blocking parameter is optimized.

\section{Simulations \label{sect:simulations}}

\subsection{SU(3) pure gauge theory \label{sect:pure_SU3}}

At the Gaussian $g=0$ FP of the pure gauge SU(3) model the gauge coupling is relevant, the
theory is asymptotically free.  According to Eq. \ref{eq:db_pert} 1-loop perturbation theory  predicts that  the bare step scaling function is constant, independent of the
gauge coupling. The 2-loop corrections are small in a wide range of coupling, $s_b^{(\rm{pert})}\approx0.59$ is a good approximation.
 The  bare step scaling function  was studied in Refs.\cite{Bowler:1984hv,Hasenfratz:1984bx,Hasenfratz:1984hx,Gupta:1984gq}
with the 2-lattice matching method. Here I repeat some of those
calculations with a different block transformation and extend them
to larger volumes and statistics. Where they overlap, the results I present below are consistent with the original calculations. 
This section mainly serves as a test of the method.

I generated 200-300 independent configurations at several coupling values 
with the Wilson plaquette gauge action 
and  calculated the bare step scaling function $s_b(\beta;s=2)$ 
 matching $32^{4}$ volumes
 on to $16^{4}$ , and also $16^{4}$ volumes to $8^{4}$.
 The $32^{4}$ volume can be blocked up
to 4 times and compared to the $16^{4}$ volume that is blocked
up to 3 times. At each blocking level I measured 5 operators: the plaquette,
the 3 6-link loops and a randomly chosen 8-link loop. 

\begin{figure}
\includegraphics[width=0.65\textwidth]{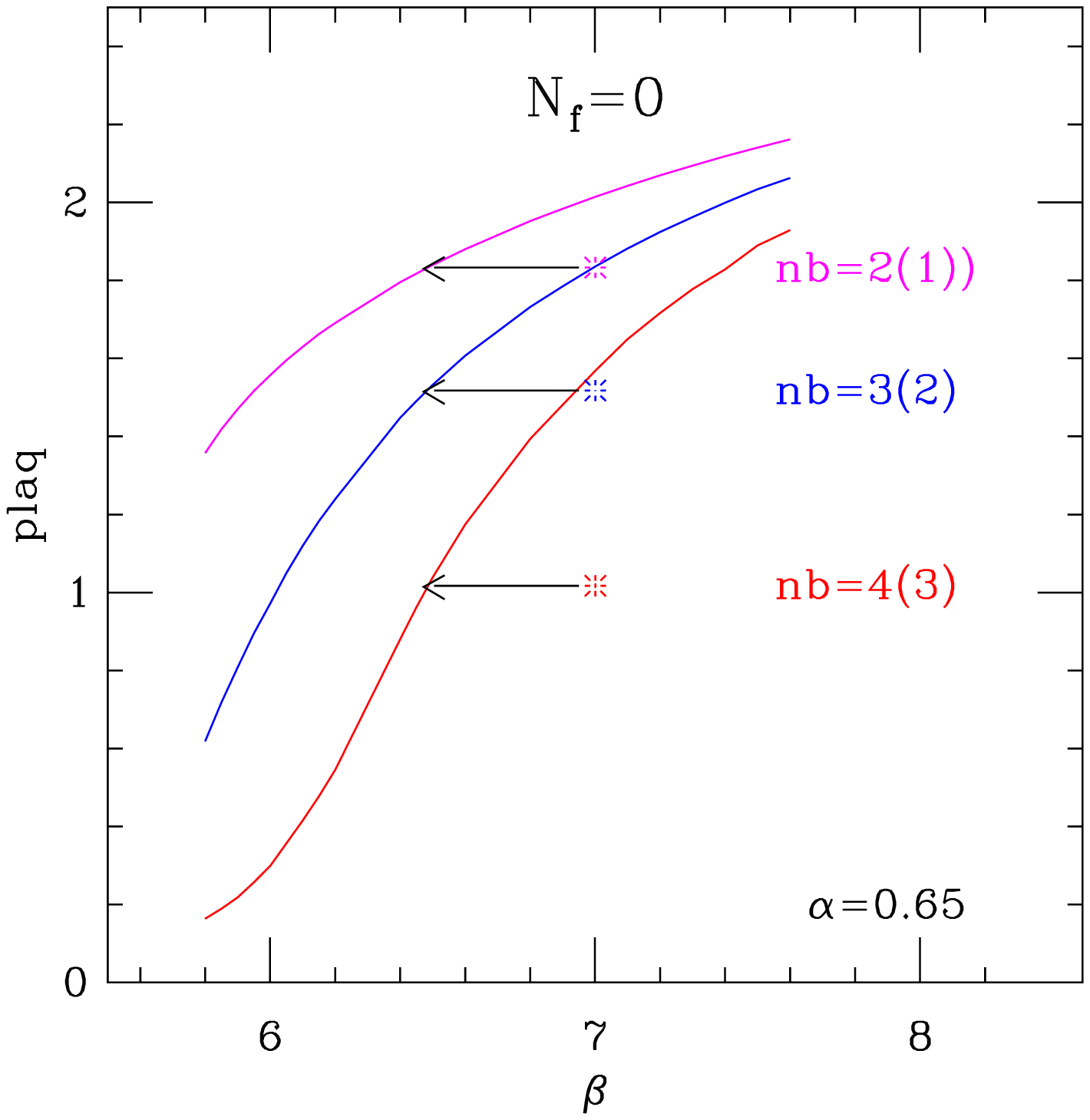}
\caption{The matching of the plaquette for pure gauge SU(3) theory. The simulations
were done on $32^{4}$ volumes at $\beta=7.0$ (symbols) and $16^{4}$
volumes at many coupling values (solid interpolating lines). The configurations
were blocked with $s=2$, $\a=0.65$ parameter block  transformation 2(1) (cyan), 3(2) (blue) and 4(3)
times (red). \label{fig:plaq_match_nf0}}
\end{figure}

Figure \ref{fig:plaq_match_nf0} illustrates the 2-lattice MCRG.
The plot  shows the matching of the plaquette 
 with the $s=2$ renormalization group transformation  of Eq. \ref{eq:block-trans} and blocking parameter   $\a=0.65$.
The  $32^{4}$ volume simulations
were done  at $\beta=7.0$, and the cyan, blue and red
symbols are the values of the blocked plaquette after 2, 3 and 4 
blocking steps. The solid curves interpolate the plaquette values, measured at many couplings on $16^4$ volumes,  after
1, 2 and 3 blocking steps. The $32^{4}$
data match the $16^{4}$ values at $\beta'=6.49$  for all blocking levels.
 The final blocked volume is   $2^{4}$, but finite size
effects are minimal as one always compares observables on the same
volume.   

\begin{figure}
\includegraphics[width=1\textwidth,clip]{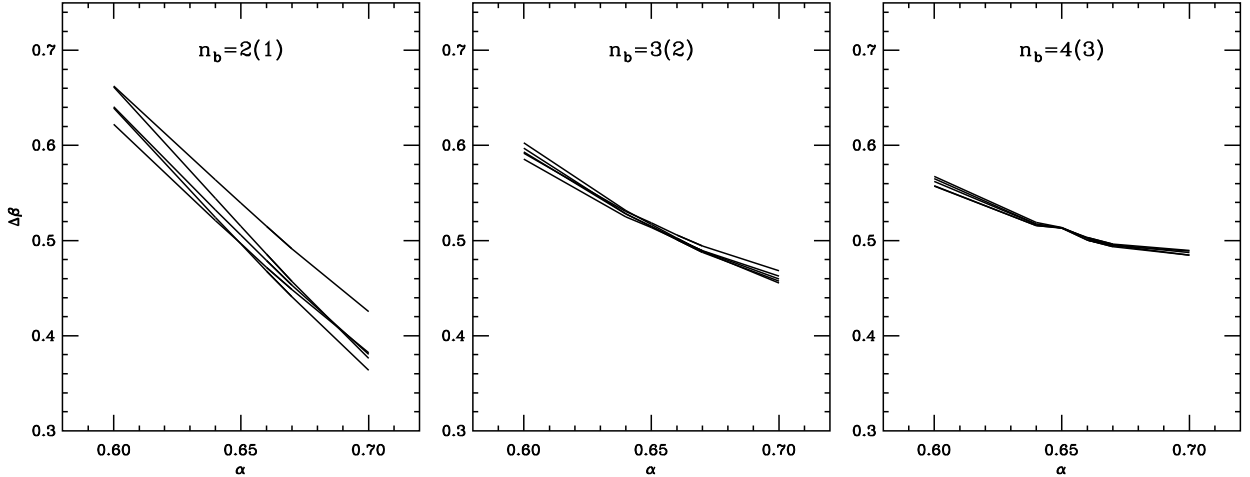}
\caption{Matching at $\beta=7.0$ from $32^4$ to $16^4$ lattices.  The matching values 
$\Delta\beta$ 
as the function of the blocking parameter $\a$ are shown for the 5 different 
operators measured. Left panel: blocking level $n_b=2(1)$; middle panel $n_b=3(2)$;
right panel $n_b=4(3)$.   }
\label{fig:sd_nf0}
\end{figure}


\begin{figure}
\includegraphics[width=0.65\textwidth,clip]{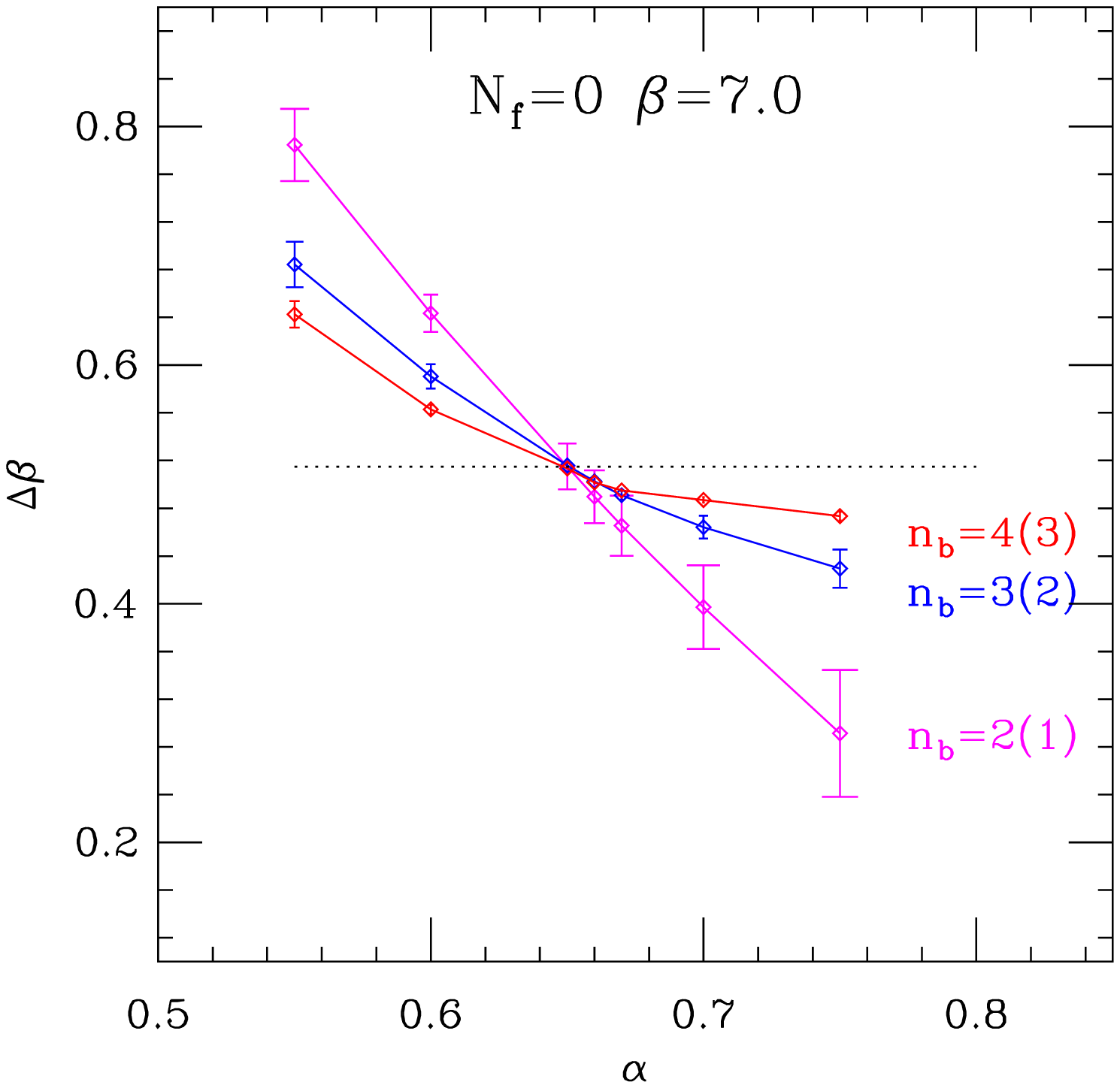}
\caption{Matching at $\beta=7.0$ from $32^4$ to $16^4$ lattices.  The average matching values 
$\Delta\beta$ 
as the function of the blocking parameter for the last 3 blocking levels are shown. Note that the "error bars"
denote the spread of the predictions for the 5 different
operators used and thus represent systematical errors. 
}
\label{fig:deltabeta_vs_alpha_b7.0_nf0}
\end{figure}

The matching can be repeated with different operators and RG transformations. Figure \ref{fig:sd_nf0} shows 
the difference between the matched couplings,
\bee
\D\b = \b -\b'\,,
\ee
as the function of the blocking parameter $\a$ for the 5 different operators at the last 3 blocking levels for $\b=7.0$.
The plots show two trends. First, the spread of the predicted $\D\b$ values from the different operators decrease with increasing blocking level, signaling that the RG flow lines are approaching the RT of the block transformation. Second, the dependence of $\D\b$ on the blocking parameter decreases with increasing blocking levels suggesting a unique value for $\D\b$ in the $n_b\to\infty$ limit.

 Figure  \ref{fig:deltabeta_vs_alpha_b7.0_nf0}
 summarizes the plots of Figure \ref{fig:sd_nf0}. The average matching $\D\b$ values are plotted as the function of $\a$ for the last three blocking levels.  The ``error
bars''  show the standard deviation (spread) of the predicted values, therefore they 
   represent  the systematic errors
of the matching procedure. The statistical errors are small, 
comparable to the systematic errors only at the last blocking level 
at the best matching around $\alpha=0.65$.
The 3 different blocking levels converge around $\alpha=0.65$, the same value where the spread
from the different operators is minimal,  predicting the relation between the lattice spacings $a(\b'=6.485) = 2 \, a(\b=7.0)$. 
In the $n_b\to\infty$ limit the quantity  $\Delta\beta(\beta)=\beta-\beta'$ 
is the bare step scaling function $s_b(\b;s=2)$, analogue to the 
step scaling function of the renormalized coupling used in the SF formalism. 
In the following I will use the intersection
of the last two blocking levels  to identify $s_b(\b)$ as it is usually less sensitive 
to the statistical errors than the spread of the  individual operators.
\begin{table}
\begin{tabular}{|c|c|c|c|c|}
\hline 
$\beta_{\hat{L}}$ & $\a_{\rm{opt}}$ &
                    $s_b,\,\hat{L}=32$ & 
                    $s_b,\,\hat{L}=32$ & 
                    $s_b,\,\hat{L}=16$  \tabularnewline
 & &  $n_{b}=3(2)$ & $n_{b}=4(3)$ & $n_{b}=3(2)$  \tabularnewline
\hline
\hline 
6.0 & 0.71 &  &  & 0.365(4)  \tabularnewline
\hline 
6.2 & 0.72 &  &  & 0.410(7)  \tabularnewline
\hline 
6.4 & 0.71 & 0.451(12) & 0.448(10) & 0.468(16) \tabularnewline
\hline 
6.6 & 0.69 & 0.488(15) & 0.483(5) & 0.496(13)  \tabularnewline
\hline 
6.8 & 0.66 &  &  & 0.511(19)  \tabularnewline
\hline 
7.0 & 0.66 & 0.517(27) & 0.515(6) & 0.516(10)  \tabularnewline
\hline 
7.2 &  0.63 & &  & 0.536(26)  \tabularnewline
\hline 
7.4 & 0.61 & 0.548(38) & 0.571(6) & 0.575(42)  \tabularnewline
\hline 
7.8 &0.60 & 0.558(34) & 0.575(5) & 0.573(42)  \tabularnewline
\hline
\end{tabular}

\caption{The bare step scaling function for the pure gauge SU(3) system.
The second column list the optimal blocking parameter.  The  third and fourth columns are results from simulations on $32^{4}$
volumes matched to $16^{4}$  after 3(2) and 4(3)
blocking steps. The last column shows results from $16^{4}$ volumes
matched to $8^{4}$ after 3(2) blocking steps.  }

\label{tab:pure_gauge}

\end{table}

 Table \ref{tab:pure_gauge} summarizes
the results  at different couplings and volumes, together with the optimal blocking parameters $\a$. The data indicate consistency 
between the different volumes and increasing blocking levels. This observation will be important 
in the study of the $N_f=4$ and 16 systems where matching is done on  $16^4\to8^4$ lattices only.
The uncertainty of the  predictions from the $16^4\to8^4$ are considerably larger than from the larger
volumes. This is not statistical, rather reflects the fact that the systematical errors of the matching after 3(2)
blocking steps are larger than after one more blocking level. It is possible that a different block transformation
would give better matching. I have not been able to modify the scale $s=2$ transformation to make it better. Adding HYP smearing\cite{Hasenfratz:2001hp} 
before constructing the blocked links pulls the RT closer, but at the same time reduces the dependence of
the expectation values on the couplings, thus increasing the  errors. It would be worthwhile to try combining the block transformation with
a  simple APE smearing, or use a variation of the scale $s=\sqrt{3}$ transformation of Refs.\cite{Patel:1984dw,Gupta:1984gq} that would allow more
blocking steps from the same lattice size.

\begin{figure}
\begin{center}
\includegraphics[width=0.65\textwidth,clip]{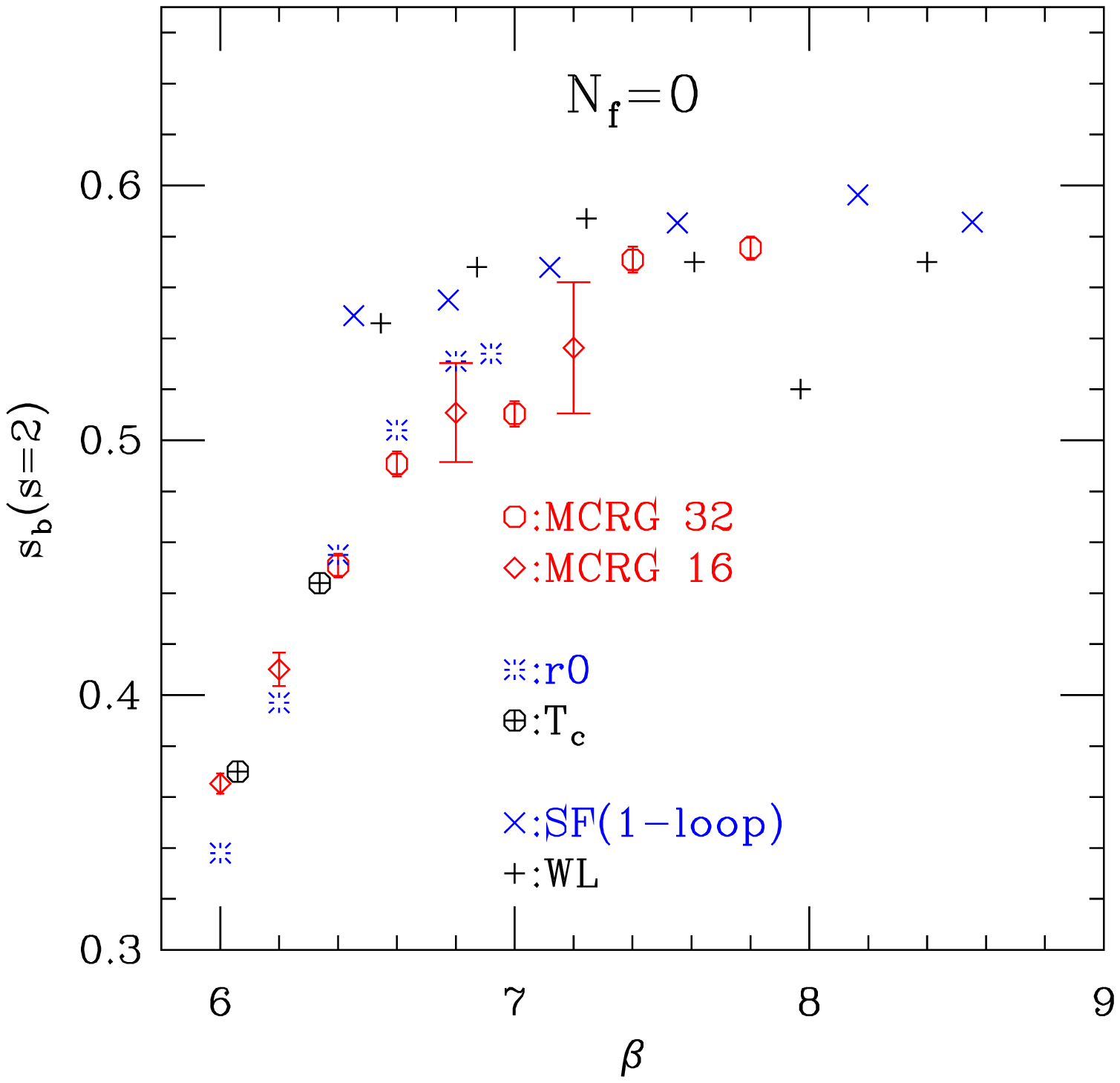}
\end{center}

\caption{The bare step scaling function $s_b(\b;s=2)$ for the pure gauge SU(3) system
as predicted by different methods. The 1-loop perturbative prediction is $s_b^{(\rm{pert})}(s=2)=0.59$. 
\label{fig:db_nf0}}

\end{figure}

The quantity $s_b(\beta;s)$ is the bare step
scaling function for scale change $s=2$. One can predict its value using physical observables  like the  Sommer
parameter $r_0$\cite{Sommer:1993ce} or the critical temperature $T_c$ . The SF calculation or the recently proposed
new method to calculate the renormalized coupling based on Wilson loop matching (WL) \cite{Bilgici:2009kh} can also predict  $s_b(\beta)$. 

In Figure \ref{fig:db_nf0} I compare the MCRG result for the bare step scaling function with predictions from
  other methods. Note that I show errors only for the MCRG results. I used the interpolating formula from Ref.
\cite{Necco:2001xg} to find $(\beta,\beta')$ pairs where $r_0/a$ differ by a factor of 2, while for $T_c$ I used 
the $N_T=8$ and 4, and $N_T=12$ and 6 transition temperatures from Ref. \cite{Boyd:1996bx}.  
In case of the SF and WL calculations I attempted to find matching $(\beta,\beta ')$ pairs  by identifying bare couplings where the renormalized SF couplings are related as $\bar g^2(\hat L) = \bar g'^2(2\hat L)$. In principle this relation should be taken in the $\hat{L} \to \infty$ limit but the numerical data do not show significant finite volume effects.
 The predictions in Figure   \ref{fig:db_nf0}  use data from the 1-loop
improved SF  \cite{Luscher:1993gh}.
 The bare couplings used in the  2-loop improved SF paper do not match close enough to use
them in this analysis \cite{Capitani:1998mq}.  

All the above calculations use the Wilson plaquette gauge action,
so in the scaling regime they should give the same prediction. It is very satisfying to see the agreement between
 MCRG, $r_0$ and $T_c$ even at relatively strong couplings. In the range $\beta \in (6.0,7.0)$ the predicted
values differ considerably from the 2-loop perturbative results.
It is difficult to measure $r_0$ or $T_c$ at much finer lattice spacings and show perturbative scaling for them.
On the other hand both
 the  SF, WL and MCRG  methods approach $s_b^{(\rm{pert})}$, but the latter one only at  $\beta\ge7.0$. The relatively large difference between the SF
and $r_0$ data was discussed and analyzed in Ref. \cite{Necco:2001xg} where it was also noted that the 2-loop improved SF 
shows significantly smaller scaling violations relative to $r_0$.

Based on the results presented in this section the 2-lattice MCRG matching method could be competitive with other methods in determining the running coupling of asymptotically free theories when it is combined with an independent definition of the renormalized coupling in the weak coupling regime.

\subsection{$N_f=4$ flavor model \label{sect:nf4}}

The 4 flavor SU(3) gauge theory is expected to be confining and chirally broken even at large gauge couplings. 
At the  Gaussian $g=0,\,m=0$ FP both the mass and the gauge coupling are relevant operators, the 
model is asymptotically free.
Perturbation theory predicts    $s_b^{(\rm{pert})}(s=2)=0.45$.
 
 The results I present here were obtained using  nHYP smeared staggered fermions \cite{Hasenfratz:2007rf}. 
 I chose nHYP smearing as it significantly reduces taste breaking of staggered fermions and therefore 
 even in strong coupling has manageable lattice artifacts.
 
 \subsubsection{The nHYP staggered action \label{sect:nHYP_action}}
 
 Very little is known about the 4-flavor system with nHYP or HYP smeared fermions. 
 The finite temperature  phase transition of the HYP smeared model
 was studied with the partial-global Monte Carlo update in Ref. \cite{Hasenfratz:2001ef}. The phase transition in the chiral limit is expected to be first order, 
 most likely extending to finite mass before turning into a crossover at large masses. 
 Simulations with thin link staggered fermions have confirmed this, finding a strong discontinuity even at fairly large quark masses.
  The  conclusion of Ref. \cite{Hasenfratz:2001ef} was quite different: we found no signal for discontinuity, the 
 phase transition appeared to be a crossover both for $N_T=4$ and 6 even at fairly
  small masses. The updating technique used in Ref. \cite{Hasenfratz:2001ef}  was not efficient enough to pursue much larger volumes, and  
   we did not continue our investigation of the $N_f=4$ system. 
   
\begin{figure}[htbp] 
   \centering
   \includegraphics[width=1\textwidth,clip]{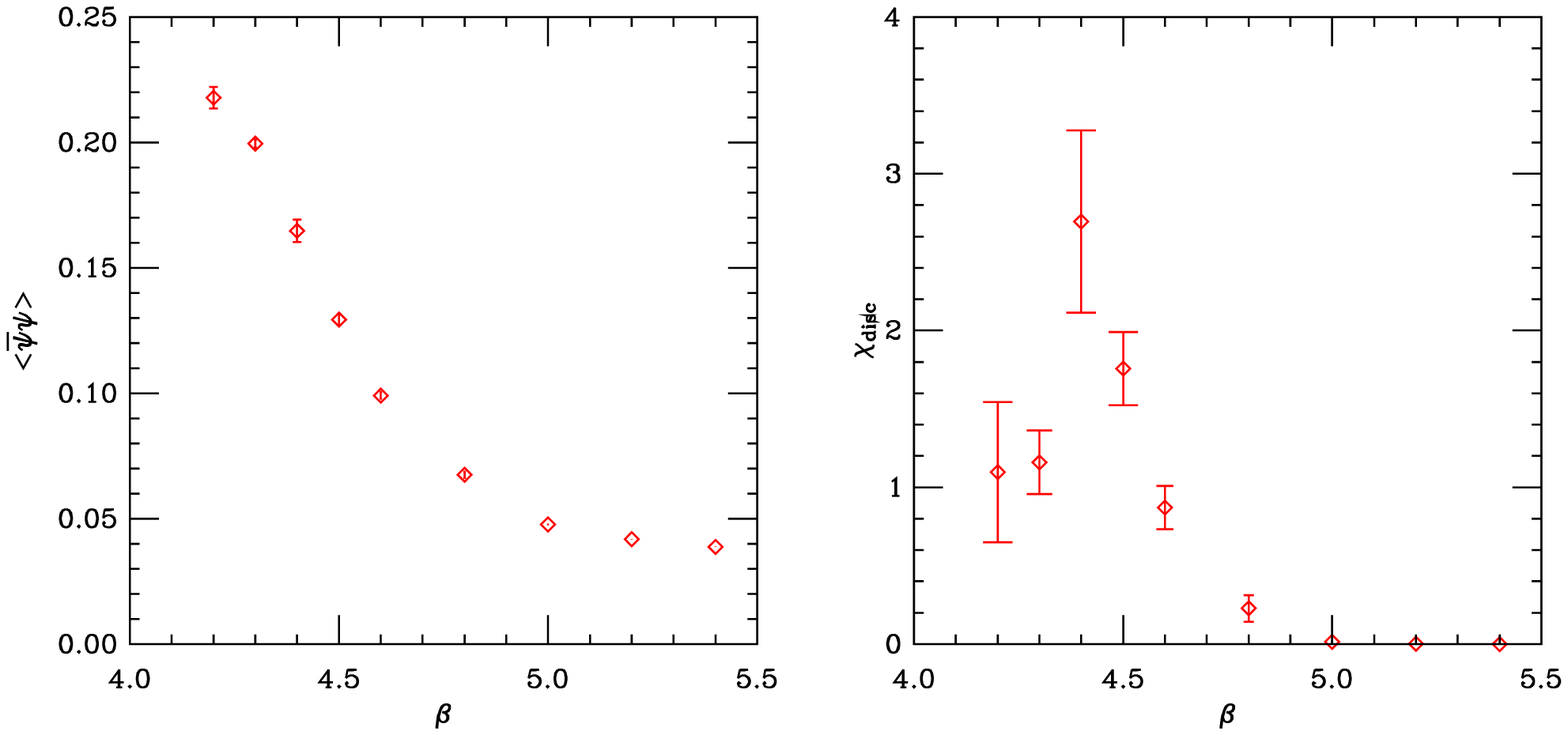} 
   \caption{ The condensate and the disconnected chiral condensate on $8^3\times4$ volumes  at $m=0.04$ in the $N_f=4$ theory.  }
   \label{fig:nf4_NT4}
\end{figure}
\begin{figure}[htbp] 
   \centering
   \includegraphics[width=1\textwidth,clip]{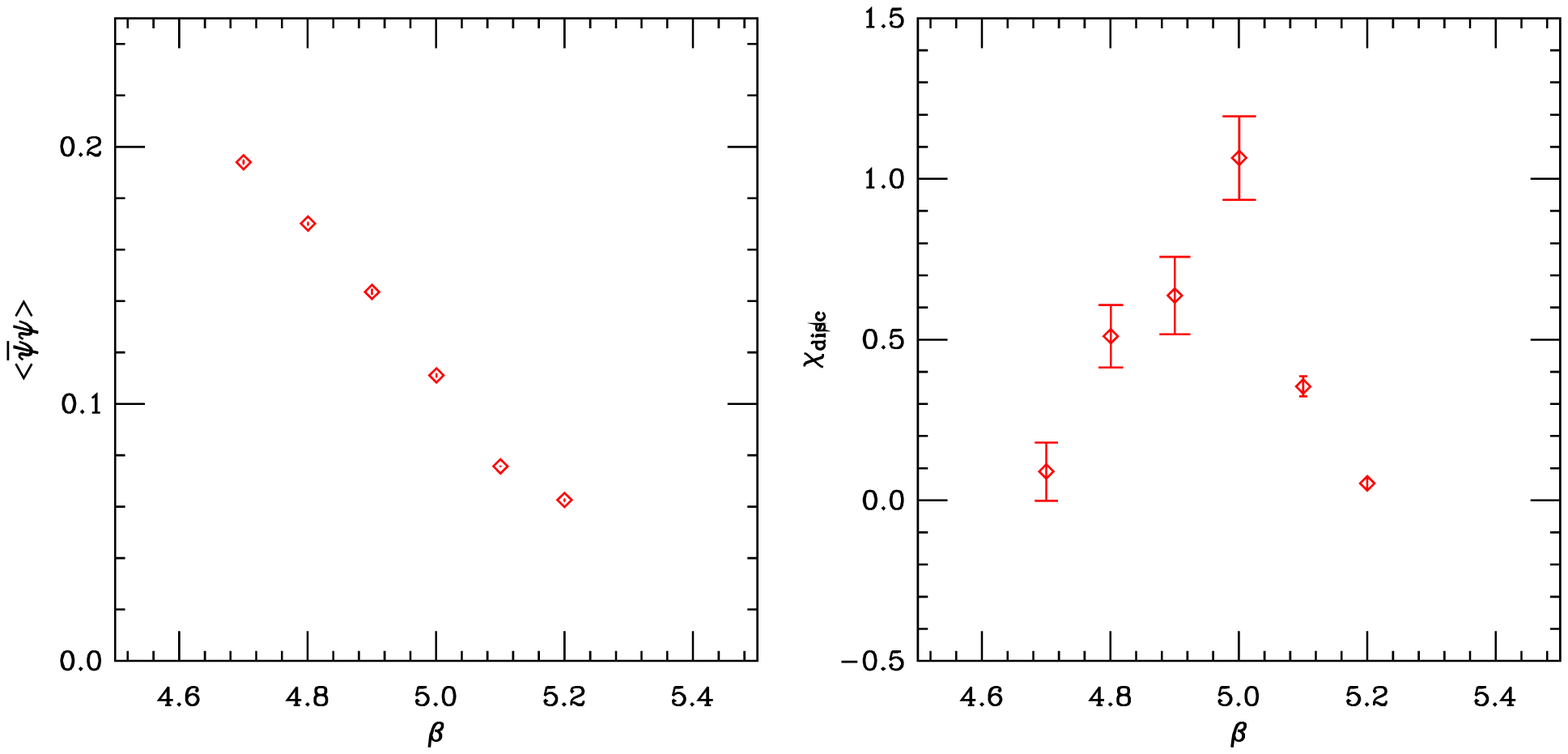} 
   \caption{ The condensate and the disconnected chiral condensate on $12^3\times6$ volumes at $m=0.05$ in the $N_f=4$ theory.  }
   \label{fig:nf4_NT6}
\end{figure}

   The nHYP smeared action is nearly identical to the HYP smeared one, but nHYP is differentiable and the efficient  molecular dynamics 
   update can be used with it \cite{Hasenfratz:2007rf}.
  I have confirmed the raw 
 data of Ref. \cite{Hasenfratz:2001ef} with the  nHYP action, and extended it further
  toward the strong coupling region. Figure \ref{fig:nf4_NT4} shows the condensate 
  $\langle \bar{\psi}\psi\rangle$ and the disconnected  chiral susceptibility \cite{Bernard:1996zw}
 \bee
  \chi_{\rm{disc}}=\langle\langle\bar{\psi}\psi\rangle^2_{\rm{conf}}\rangle_U - \langle\langle\bar{\psi}\psi\rangle_{\rm{conf}}\rangle^2_U\,
  \ee
  on $8^3\times4$ lattices at $m=0.04$. The condensate is almost identical to 
  Figure 5 of \cite{Hasenfratz:2001ef}, a smooth function of the gauge 
  coupling with no obvious sign of discontinuity. The disconnected  chiral susceptibility has a strong peak at $\b=4.4$,  
  signaling a crossover. In contrast, 
  the  data with thin link staggered fermions at $N_T=4$ show a discontinuity in the condensate $\D\langle \bar{\psi}\psi\rangle\approx 0.2$. Figure
  \ref{fig:nf4_NT6} is the same as Figure \ref{fig:nf4_NT4} but on $12^3\times 6$ lattices at $m=0.05$. Again, the condensate is smooth, the susceptibility suggests a crossover around $\b=5.0$. Obviously much more work is needed to determine the transition temperatures and the order of the phase transition accurately. Larger spatial volumes might sharpen the transition, but in any case the endpoint of the first order line  
  with nHYP fermions occurs at much smaller masses than with the thin link action.
  
  To set the scale I measured the static potential on $16^4$ lattices used later in the MCRG study. I found $r_0/a = 5.8(3)$ at $\b=5.4$, $m=0.01$.  Ref.   \cite{Hasenfratz:2001ef} quotes  $r_0/a=3.34(7)$ at $\b=5.2$, $m=0.04$ and $r_0/a=2.2(1)$  at $\b=5.0$, $m=0.10$. 

\subsubsection{MCRG matching}

\begin{figure}[htbp] 
   \centering
   \includegraphics[width=0.65\textwidth,clip]{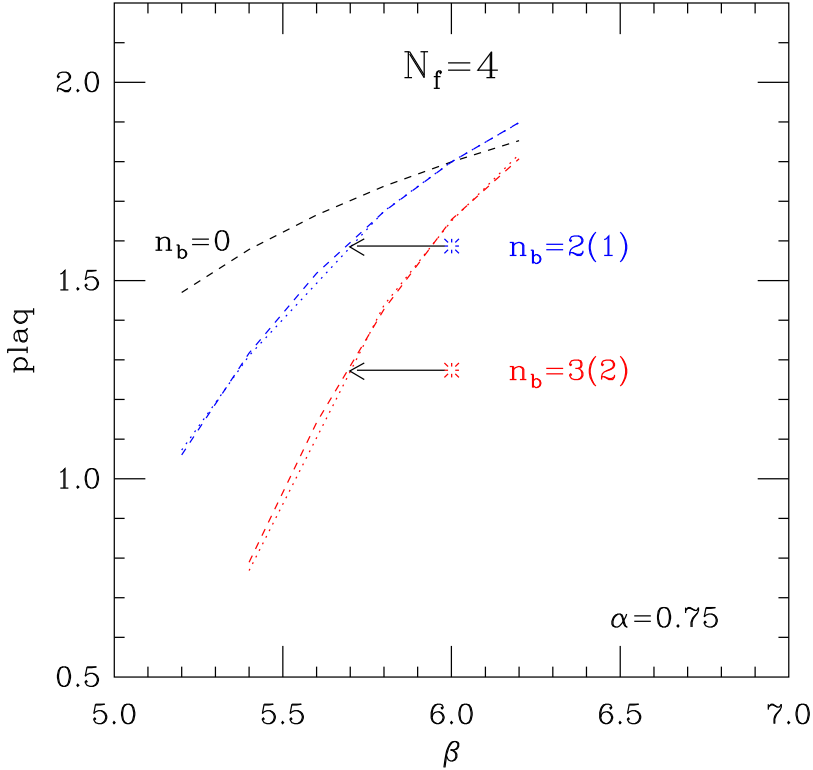} 
   \caption{Matching of the plaquette in the $N_f=4$ model. The  blue and red symbols show the 
plaquette value after 2 and 3 blocking steps on the $16^4$ volumes at 
$\b=6.0$, $m=0.01$. The  lines interpolate the plaquette  
on the $8^4$ volumes at several couplings and $m=0.015$ (dashed) and $m=0.025$ (dotted) 
after 1 and 2 blocking levels. The black line is the unblocked plaquette on the $8^4$ lattices.
The  dashed and dotted curves are barely distinguishable. }
   \label{fig:plaq_match_nf4}
\end{figure}

In principle  MCRG matching could be done  similarly to the pure gauge SU(3) model,
but since the fermionic model has 2 relevant operators, the matching requires tuning in both the gauge coupling and the mass .
This is a considerably harder numerical task than a single parameter matching. One can reduce this 
complication by setting one of the relevant couplings to its critical value, since then only the other
coupling has to be matched. The critical value of the mass is $m=0$. Simulations in small  volumes are 
possible even with vanishing mass. In addition the dependence of the local observables used in the matching 
is so weak on the mass that a small mass in the simulations is also acceptable. Setting the mass to zero or to a small value allows
matching in the gauge coupling only and the bare step scaling function  can be calculated in the same way
as for the pure gauge system.

 To calculate the step scaling function   I considered $16^4\to8^4$ matching at several 
gauge couplings between $\b\in(5.4,8.0)$ (see Table \ref{tab:NF4}). All the $16^4$  configurations except $\b=5.4$  are in the deconfined, chirally symmetric  phase, but that does not matter for the MCRG matching method.
 I have generated 100-150 configurations on the larger volumes and $\approx300$ 
configurations on the smaller ones, separated by 10 molecular dynamics trajectories. I have used the 
same 5 operators in the matching as in the pure gauge SU(3) system.

\begin{figure}
\includegraphics[width=0.65\textwidth,clip]{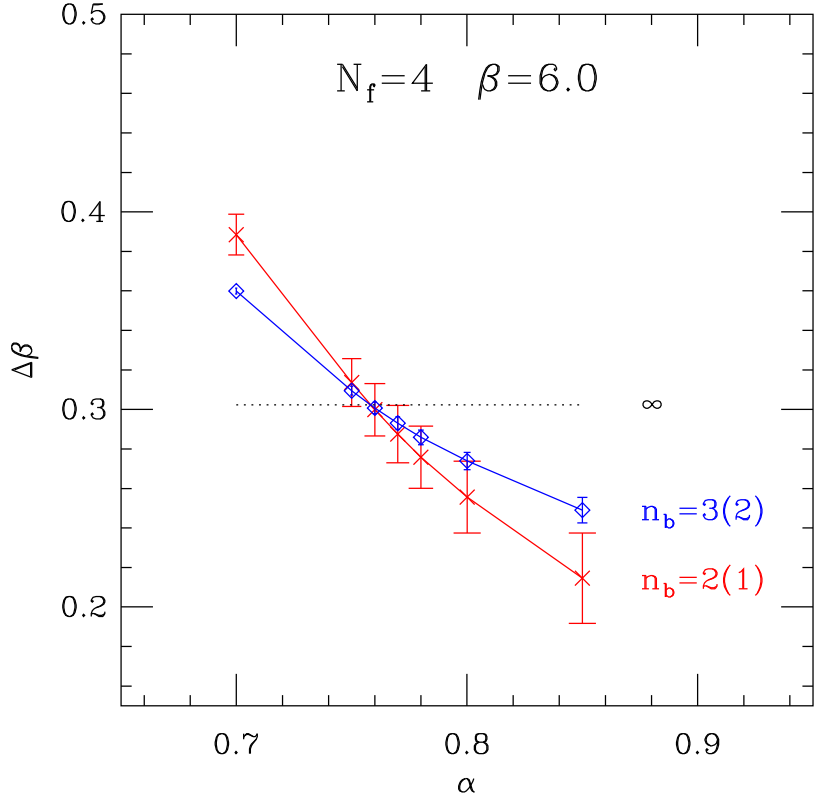}
\caption{Matching at $\beta=6.0$ from $16^4$ to $8^4$ lattices in the $N_f=4$ model.  The matching values 
$\Delta\beta$ 
as the function of the blocking parameter for the last 2 blocking levels are shown. Note that the error bars
denote the spread of the predictions for the 5 different
operators used and thus represent systematical errors. 
}
\label{fig:deltabeta_nf4}
\end{figure}

\begin{table}

\begin{tabular}{|c|c|c|}
\hline 
$\beta_{16}$ & $\a_{\rm{opt}}$  & $s_b(s=2)$ \tabularnewline
\hline
\hline 
5.4 &  0.81 &0.388(10) \tabularnewline
\hline 
5.6 &  0.81 &0.330(13) \tabularnewline
\hline 
5.8 &  0.77 &0.335(20) \tabularnewline
\hline 
6.0 &  0.76 & 0.303(25) \tabularnewline
\hline 
6.4 & 0.67 & 0.400(29)  \tabularnewline
\hline 
6.8 &  0.67 & 0.365(42) \tabularnewline
\hline
7.2 &  0.63 & 0.404(39) \tabularnewline
\hline
8.0 & 0.59 &0.470(53)  \tabularnewline
\hline
\end{tabular}

\caption{The bare step scaling function 
 $s_b(s=2)$ for the 4-flavor simulation. The second column lists the optimal blocking parameter $\a$.\label{tab:Nf4}}

\label{tab:NF4}
\end{table}

\begin{figure}[htbp] 
   \centering
   \includegraphics[width=0.65\textwidth,clip]{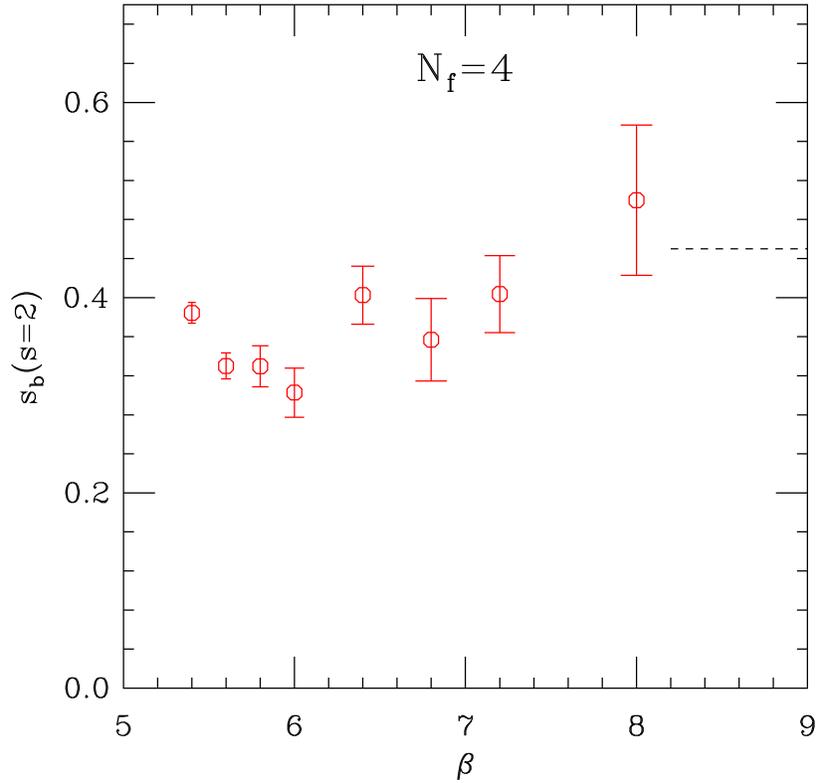} 
   \caption{The  bare step scaling function for 4 flavor SU(3) theory. The dashed line indicates the 1-loop perturbative value.}
   \label{fig:db_nf4}
\end{figure}

On the $16^4$ lattices I chose $m=0.01$.  If the critical exponent for the mass were its engineering dimension, the matching
mass on the smaller volume would be $m=0.02$. I generated $8^4$ lattices  with  $m=0.015$ and 0.025 to bracket this value and  
 to check for any dependence on the mass.  The matching for the plaquette is shown in Figure \ref{fig:plaq_match_nf4}. The 
  blue and red symbols represent the plaquette value after 2 and 3 blocking steps on the $16^4$ lattice at $\b=6.0$ while the dashed and dotted lines interpolate the 1 and 2 times blocked plaquette values on the $8^4$ lattices with $m=0.015$ and $m=0.025$.  For completeness I also  show the unblocked plaquette on the $8^4$ volumes (black line) though there is no  consistent matching at this level.
The dotted and dashed  curves are barely distinguishable. None of the other observables show any dependence on the mass even after 3
blocking steps beyond the  fairly small statistical errors of the simulations, implying that the system indeed can be considered 
critical in the mass. 

 The dependence of the matched values on  the blocking parameter $\a$ is similar to the pure gauge case. The analogue of Figure \ref{fig:deltabeta_vs_alpha_b7.0_nf0} is Figure \ref{fig:deltabeta_nf4}. Since for  $16^4\to8^4$ matching only two blocking levels can be used, one has only two sets of predictions, $n_b=2(1)$ and $n_b=3(2)$.  In Sect. \ref{sect:pure_SU3}, Table \ref{tab:pure_gauge}  I found that  one 
can safely identify the optimal blocking parameter  and matching value from the intersection of 
these blocking levels even on $16^4\to8^4$ matching.

Table \ref{tab:NF4}  and Figure  \ref{fig:db_nf4} summarize the simulation results. 
The step scaling function $s_b(s=2)$ is consistent with asymptotic freedom and approaches the 2-loop perturbative prediction
at weak couplings.
Just like in the pure gauge system, the data in Table \ref{tab:NF4}  could be used to determine the running coupling of the
4-flavor model. To do so one needs to calculate a renormalized coupling at $\b=8.0$ and connect it through perturbation
theory to a continuum scheme. The change of the lattice scale between $\b=5.4$ and $\b=8.0$ can be determined form the data, while
the lattice spacing can be obtained by measuring a physical quantity at some strong coupling.  Improved block transformation or larger volumes
would allow a more precise determination of $s_b(\b)$. Higher statistics especially at the larger $\b$ values, would also help.

\subsection{$N_{f}=16$ flavor model \label{sect:nf16}}

The  16 flavor SU(3) model is still  asymptotically free, but 
 2-loop perturbation theory predicts the emergence of an IRFP
at weak  gauge coupling \cite{Banks:1981nn}. Continuum limits can be defined both at the Gaussian 
 FP and at the new IRFP. In the former case both the gauge coupling and the mass has to be tuned
towards the FP, in the latter one the gauge coupling is irrelevant, only the mass has to be tuned to $m=0$. 
There is  no confinement or spontaneous chiral symmetry breaking 
 in the  weak  coupling phase, the continuum massless theory at the IRFP is conformal. The conformal phase was first identified in Ref. \cite{Heller:1997vh}.
 
It is generally believed that  in the strong coupling the lattice model is confining and chirally
broken, so there has to be a (bulk) transition separating the strong
and weak coupling phases. 
Since the bulk transition is a lattice artifact, it is probably not
  associated with  critical behavior or continuum quantum field theory.
Most likely it is  a first order phase transition at $m=0$ and it might  extend to $m>0$ before 
turning into a crossover \cite{Damgaard:1997ut,DeGrand:2009mt}. One should mention that some numerical results indicate that 
this confining phase might not even exists \cite{Iwasaki:2003de}. 

\subsubsection{MCRG around an infrared fixed point}

Assuming that the  simulations are done in the conformal phase, the  behavior 
 we expect from MCRG depends on whether we study  the critical ($m=0$) or the   $m\ne 0$ phases:

\begin{itemize}
\item 
On the $m=0$ critical surface at very weak coupling the system is in the attractive region of the 
Gaussian FP.  2-lattice matching could reveal the running of the gauge coupling,  $s_b^{(\rm{pert})}=0.016$ 
 from the 1-loop $\beta$ function, though
numerical simulations probably will not be close enough to the Gaussian FP to see this behavior. It is much more likely
that the RG flow will be determined by the Banks-Zaks IRFP. At this FP all operators are irrelevant when $m=0$. The 2-lattice matching is designed 
to map out the flow speed in the relevant direction, so we have to re-evaluate the method when there is no relevant operator.

When all  operators are irrelevant, they all flow into the FP according to their scaling dimensions. In the $n_b\to\infty$ limit 
all expectation values approach their FP value independent of the bare couplings, so matching is meaningless.  At finite $n_b$ 
the RG flow can pick up  the "least irrelevant" operator.  If there is one operator with a nearly zero
scaling dimension matching can make sense: after a few RG steps all other operators are already in the FP, so the flow line
follows that single operator.  According to Eq. \ref{eq:gamma_g} the scaling 
exponent of the gauge coupling for the $N_f=16$ flavor theory at the Banks-Zaks FP  is small, $\gamma_g= -0.01$, it  
is almost a marginal operator. The scaling exponents of the other irrelevant operators are likely close to their engineering dimensions, starting 
at  $\gamma\approx-2$, so these operators will die out much faster than  the gauge coupling.  

The picture we expect in the 2-lattice matching is now clear. Since the gauge coupling is nearly marginal, the matching will follow its flow.  $s_b(\b)$  is given by 
Eq. \ref{eq:db_IRFP} and for a marginal operator $s_b(\b)=0$ near the FP. 
For an almost marginal operator  $\D\b=\b-\b'$
can be  either positive or negative, depending on whether the FP of the actual RG transformation is at smaller or larger gauge coupling. 

The evolution of the blocked operators can also signal the IRFP.  As the RG flow approaches the IRFP all expectation values approach their 
IRFP value. On the other hand if the GFP controls the system the flow lines follow the RT and run into the trivial $\b=0$ FP where all local expectation values vanish. This difference is one of the strongest signal for the existence of an IRFP in the MCRG method.

\item
At finite mass  the RG transformation is  dominated by the flow of the relevant mass operator. However the nearly marginal gauge coupling can still have a strong influence on the flow, the situation is more like  matching 2 relevant operators than matching  a single one.
The easiest way to deal with this is to set the gauge coupling to its FP value (i.e. to the value that corresponds to the IRFP of the RG transformation used) and match in the mass only. This matching  predicts  the scaling dimension (or critical exponent) of the mass.
While the IRFP of the RG transformation depends on the blocking parameter, the exponent itself is independent of both $\a$ and the gauge coupling $\b$.
\end{itemize}

\subsubsection{Numerical simulations and results}

 I concentrate on the renormalization group properties of the model 
  in the massless limit and show only preliminary results at finite mass.
  I did 2-lattice matching on $16^4\to8^4$ lattices using nHYP smeared staggered fermions \cite{Hasenfratz:2007rf}, and as in the $N_f=4$ case, the simulations were done with a small mass, $m=0.01$ on the   $16^4$ and $m=0.02$ on the $8^4$ lattices.   I have collected 100-150 configurations on the larger volumes, $\sim200$ on the smaller ones, separated by 10 molecular dynamics steps. 

On the $8^4$ lattices I  covered the coupling range $\b\in(2.4,8)$, trying to identify the bulk transition to the strong coupling  phase.  The data show no sign of a  phase transition, though for $\b<4.0$ 
the condensate starts  increasing slowly, suggesting the development  of spontaneous chiral symmetry breaking.  It is likely that the bulk transition exists only at very small, possibly vanishing, mass. This
does not contradict the results of Ref. \cite{Damgaard:1997ut} where a strong first order bulk transition was observed with  
$N_f=16$ fermions. 
As we learned from the finite temperature investigation with $N_f=4$  flavors in Sect. \ref{sect:nHYP_action},
smearing the fermionic action can soften or wash away first order phase transitions, and that might be the situation here as well.

\begin{figure}
\begin{center}
\includegraphics[width=0.65\textwidth,clip]{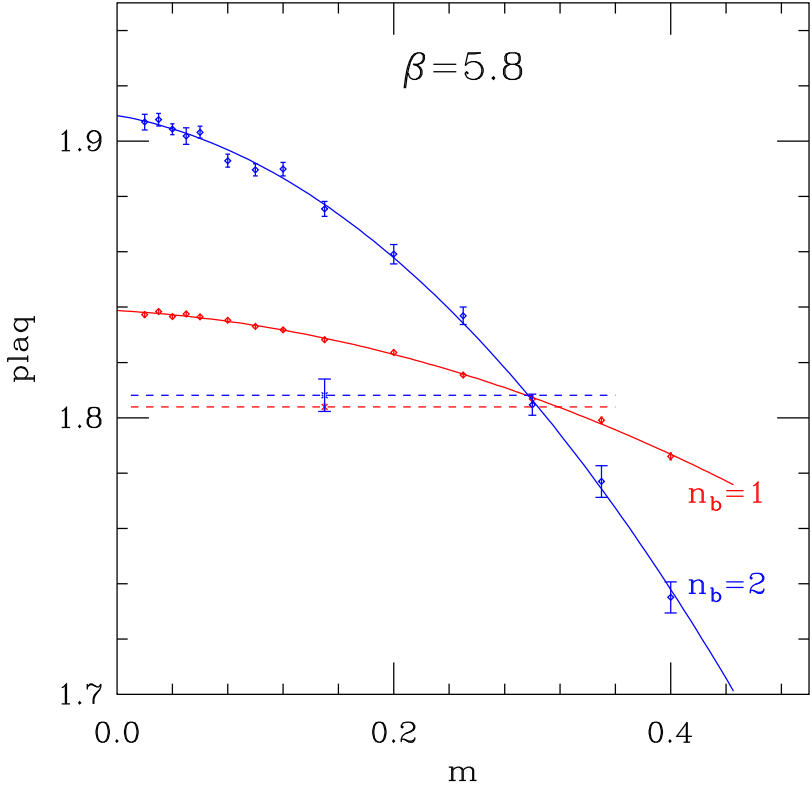}
\end{center}

\caption{The dependence of the plaquette on the mass after 1 (red diamonds and solid line) and 2  (blue diamonds and solid line)
blocking steps on $8^4$ volumes at $\b=5.8$. The bursts at $m=0.15$  show the plaquette after 2 and 3 blocking steps starting form $16^4$ volumes at the same $\b$ value. The dashed lines indicate matching in the mass.  
The blocking parameter is $\a=0.75$, close to the optimal value at $\b=5.8$. }
\label{fig:plaq_vs_mass}

\end{figure}

\begin{figure}
\begin{center}
\includegraphics[width=0.65\textwidth,clip]{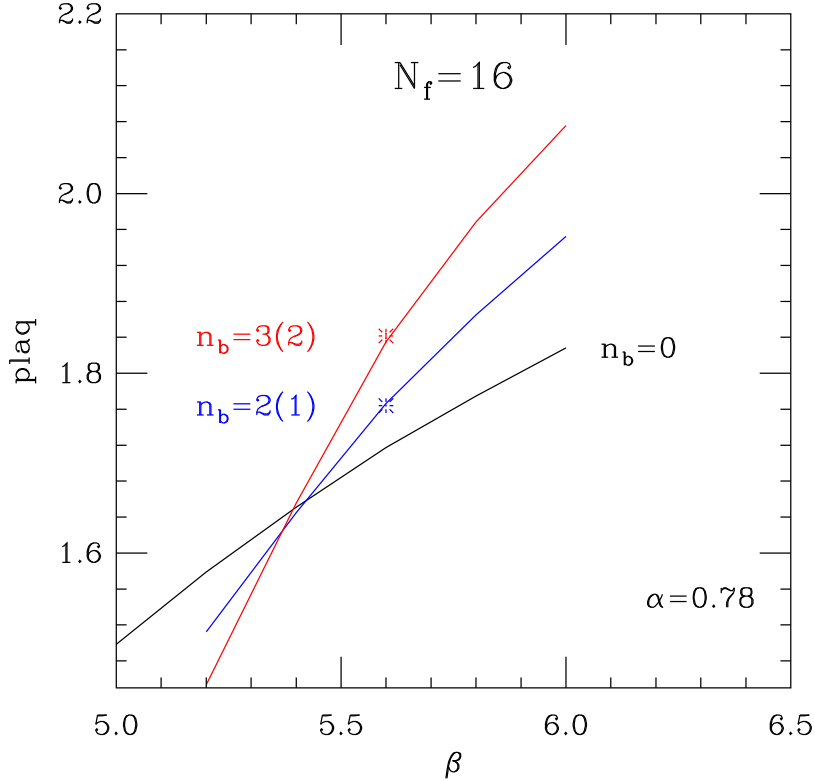}
\end{center}

\caption{The matching of the plaquette for $N_f=16$ flavor. The individual 
data points are  at $\beta=5.6$  on  $16^4$ lattices after 2 and 3 levels of blocking. 
They are compared to the  1 and 2 times blocked values as measured on the $8^4$ lattices 
(solid lines).
The matching values $\D\b\approx 0$ indicate a
nearly marginal flow. Note that the plaquette increases with $n_b$ implying that the RG flow is to an IRFP and not the $\b=0$ trivial FP.  }
\label{fig:nf16_plaq_match}

\end{figure}

Just like in the $N_f=4$ case the mass dependence of the measured operators is weak for small $m$. 
 Figure \ref{fig:plaq_vs_mass}  shows  the plaquette blocked 1 and 2 times on the $8^4$ configurations at $\b=5.8$ at different masses.
 The solid lines are simple spline extrapolations. Within errors there is no mass dependence up to about $m\le0.05$ for the plaquette.
 Other observables are similar, so for the MCRG matching the $m=0.01-0.02$ data set can be considered critical in the mass.
 
Figure \ref{fig:nf16_plaq_match} is the analogue of Figures  \ref{fig:plaq_match_nf0} and \ref{fig:plaq_match_nf4},  showing the 
matching of the plaquette. The data points are  at $\beta=5.6$  on  $16^4$ lattices after 2 and 3 levels of blocking. 
They are compared to the  1 and 2 times blocked values as measured on the $8^4$ lattices. The block transformation is 
with parameter 
$\a=0.78$, optimal for $\b=5.6$  and  matching is consistent for both blocking levels with $\D\b= 0$.

From the matching value alone it is not possible to distinguish a marginally relevant flow (like at the GFP) from an 
almost marginal irrelevant flow (expected at the IRFP).  Comparing Figure \ref{fig:nf16_plaq_match} 
to the $N_f=4$ or pure gauge Figures \ref{fig:plaq_match_nf0}  and \ref{fig:plaq_match_nf4} 
reveals an important difference. When the flow is governed
by the GFP the flow lines follow the RT towards the trivial $\b=0$ FP. In the $n_b\to\infty$ limit all expectation values vanish.  In   Figures \ref{fig:plaq_match_nf0} and \ref{fig:plaq_match_nf4} the plaquette indeed decreases with increasing blocking levels. 
On the other hand when the flow
lines approach an IRFP  all expectation values take the value at the FP. The plaquette in Figure \ref{fig:nf16_plaq_match} increases with
increasing blocking steps for $\b \ge 5.4$, indicating that the flow lines are  not running towards the $\b=0$ FP.  Eventually all  expectation values should  become independent of the blocking steps and $\b$. This trend is not obvious from the figure for  two main reasons. The first is 
that  finite volume effects are considerable  on $2^4$ lattices, the second that
with only 2 or 3 blocking steps
one approaches the FP only if the RG transformation is optimized. Nevertheless the fact that with optimal blocking the blocked operators increase with $n_b$ already signals 
that the flow runs toward an IRFP.

Other operators show similar behavior at $\b\ge 5.6$ but the results are quite different at stronger gauge couplings.  Figure \ref{fig:nf16_plaq_match} shows that the blocked plaquette values cross around $\b=5.4$ and they decreases with $n_b$  for $\b<5.4$. The location of the crossing depends on the blocking parameter but for the $N_f=16$ flavor model it never drops below $\b=5.4$.  It appears that  the flow is running towards the $\b=0$ FP  when $\b<5.4$ and to the IRFP when $\b>5.4$. 
\begin{figure}
\begin{center}
\includegraphics[width=1\textwidth,clip]{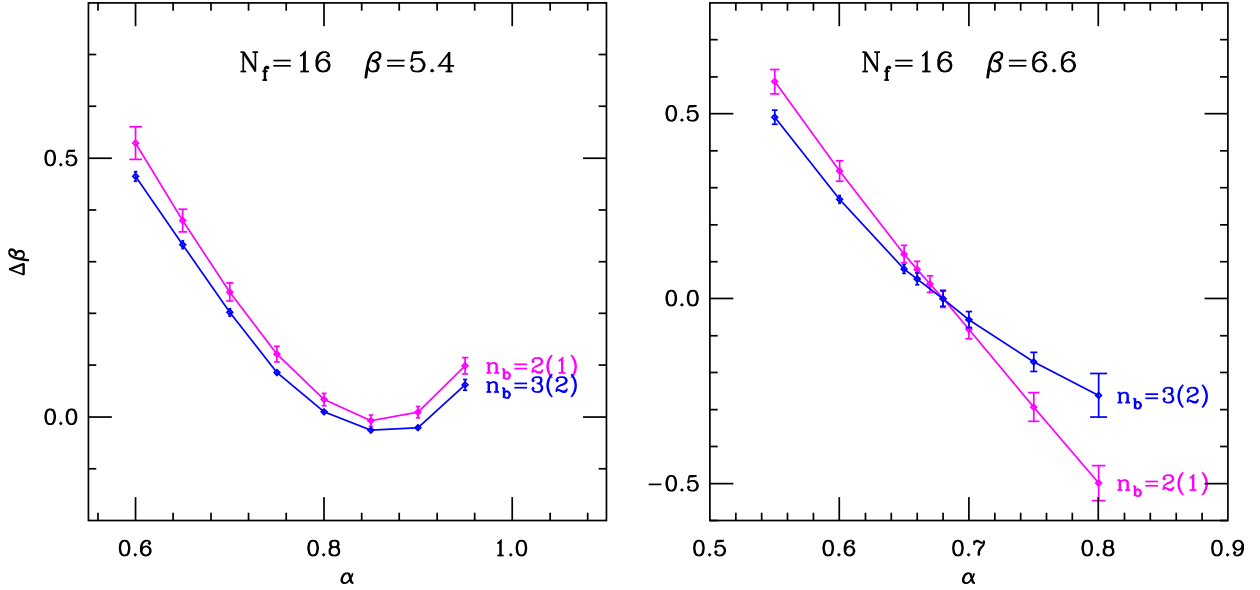}
\end{center}

\caption{Matching at $\beta=5.4$ (left panel) and $\b=6.6$ (right panel) from $16^4$ to $8^4$ lattices in the $N_f=16$ flavor theory. }
\label{fig:nf16_db_a}

\end{figure}




%
\begin{table}
\begin{tabular}{|c|c|c|}
\hline 
$\beta_{16}$ & $\a_{\rm{opt}}$ &  $s_b(s=2)$ \tabularnewline
\hline
\hline 
5.4 & & none \tabularnewline
\hline 
5.6 &  0.78 & 0.00(3)  \tabularnewline
\hline 
5.8 &  0.78 & -0.12(4)  \tabularnewline
\hline 
6.2 & 0.66 & 0.09(5)  \tabularnewline
\hline 
6.6 & 0.67 & -0.02(4)  \tabularnewline

\hline
\end{tabular}

\caption{The parameters and MCRG results of the $N_{f}=16$ flavor simulations. The second column lists the optimal blocking parameter $\a$. \label{tab:nf16}}

\end{table}

\begin{figure}
\begin{center}
\includegraphics[width=0.65\textwidth,clip]{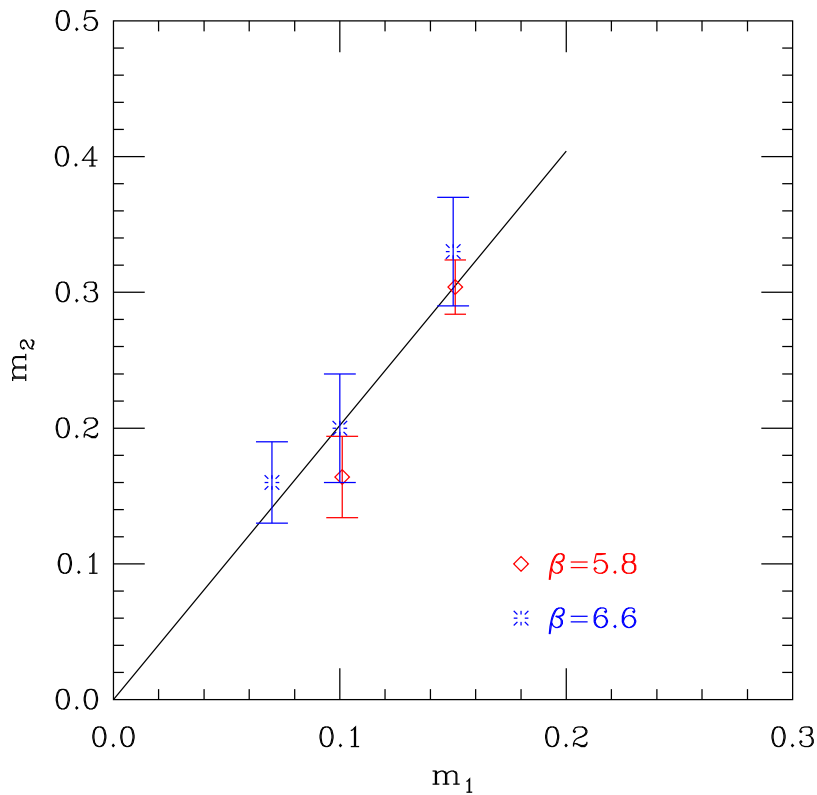}
\end{center}

\caption{Matched $(m_1,m_2)$  pairs in the $N_f=16$ system. Red diamond: $\b=5.8$, blue bursts: $\b=6.6$. The linear fit predicts
$y_m=1.02(7)$. }
\label{fig:m1_vs_m2}

\end{figure}

%
\begin{table}
\begin{tabular}{|c|c|c|c|}
\hline 
$\beta$ & $\a_{\rm{opt}}$ &  $m_1$ & $m_2$  \tabularnewline
\hline
\hline 

5.8 &  0.75 & 0.10 & 0.17(3)  \tabularnewline
\hline 
5.8 &  0.75 & 0.15 & 0.31(2)  \tabularnewline
\hline 
6.6 &0.68 & 0.07 & 0.16(3)  \tabularnewline
\hline 
6.6 & 0.68 & 0.10 & 0.20(4)  \tabularnewline
\hline 
6.6 & 0.68 & 0.15 & 0.33(4)  \tabularnewline

\hline
\end{tabular}

\caption{Matched mass pairs in the $N_f=16$ system. \label{tab:mass_match}}

\end{table}

The 2-lattice matching is also different below and above $\b=5.4$.  Figure \ref{fig:nf16_db_a} shows $\D\b$ as the function of the blocking parameter $\a$ for $\b=5.4$ and $\b=6.6$. This figure is the analogue of  Figure \ref{fig:deltabeta_vs_alpha_b7.0_nf0}. As the left panel  of  Figure \ref{fig:nf16_db_a}  shows the $n_b=2(1)$ and 3(2) blocking levels 
get close but do not actually converge at $\b=5.4$, there is no consistent matching. The situation is similar, even more enhanced, at stronger couplings. 
I have not observed this kind of behavior either with $N_f=0$ or 4, though I had simulations at even stronger couplings 
there. 
While it is
possible that higher blocking levels would predict matching and the expectation values eventually approach their FP value, 
it is more likely that  we see  the effect of a nearby bulk transition and the strong coupling confining region beyond it. 
Apparently  $\b\le5.4$ is not governed by the IRFP. The situation is entirely different at $\b=6.6$ where   predictions from the two different blocking levels converge around $\a=0.675$ predicting $\D\b=-0.022(44)$ (right panel  of  Figure \ref{fig:nf16_db_a}). Results are similar at other $\b > 5.4$ couplings, as summarized in Table \ref{tab:nf16}.  The optimal block transformation predicts $\D\b\approx0$ for all coupling values, in agreement with the expectations, i.e. that the RG flows are governed by an almost marginal operator. Combining this with the observation that with optimal blocking parameter the expectation values of the blocked operators increase leads to the conclusion that the 
$\b\in(5.6,6.6)$ coupling range in the massless limit is governed by an IRFP. 

The mass  is a relevant operator at this IRFP, therefore it should scale according to Eq. \ref{eq:mass_scale} under an $s=2$ RG transformation. 
One can calculate the exponent $y_m$  by identifying matched $(m_1,m_2)$ mass values. The gauge coupling is irrelevant, any
$\b$ in the attractive basin of the IRFP could be chosen for this. Since the gauge coupling is nearly marginal it is best to use the same  coupling at both mass values.
 For matching one can use the same operators as before or add others that are more sensitive to the ferminons. I believe that direct
fermionic observables would allow more precise matching at smaller quark masses, but as Figure  \ref{fig:plaq_vs_mass} illustrates the gauge observables also work. In fact Figure \ref{fig:plaq_vs_mass} already shows a matched mass pair. At $\beta=5.8$ with $\a=0.75$ RG transformation the $m_1=0.15$ mass on the $16^4$ configurations match  $m_2=0.318(6)$ on the $8^4$ configurations after 2(1) blocking steps, while the matching mass is $m_2=0.297(10)$ after 3(2) blocking steps. These values are for the plaquette but other observables give similar values, predicting $m_2=0.31(3)$ at the optimal $\a=0.75$ parameter. I have preliminary data at $\b=5.8$ and 6.6 at a couple of mass values. The matching $(m_1,m_2)$ pairs are listed in Table \ref{tab:mass_match}. The fit according to Eq. \ref{eq:mass_scale} predicts $y_m=1.02(7) $ as shown in Figure \ref{fig:m1_vs_m2} 
The scaling dimension of the mass is very close, within errors undistinguishable, form it engineering dimension. This is not unexpected as the IRFP is at weak coupling. In a recent publication $y_m=1.5$ was predicted for sextet fermions \cite{DeGrand:2009et}.  
It would be interesting to study the sextet model,  or the  $N_f=12$ model, where the $y_m$ might be significantly different from 1.  

\section{Conclusion} 
Renormalization group methods have been designed to study the critical behavior of statistical systems. In this paper I have shown that
they are equally suitable to study the renormalization group structure of  quantum field theories. I have used a numerical Monte Carlo Renormalization Group method to calculate the 
step scaling function and critical exponent of SU(3) gauge theories with $N_f=0$, 4 and 16 flavors. I chose these fairly well understood systems
as my goal was to test the method before using it in more relevant simulations. The paper is fairly pedagogical, explaining in detail the 2-lattice matching MCRG method.

In the $N_f=0$ case I demonstrated that the bare step scaling function predicted by the 2-lattice matching method is consistent with the more traditional 
Schrodinger functional results. It is also consistent with results obtained from the scaling of the $r_0$ parameter and  the  finite temperature phase transition 
even at strong gauge coupling, suggesting scaling, though not 2-loop perturbative scaling there. 

The $N_f=4$ and 16 flavor simulations were done using nHYP smeared staggered fermions. I chose nHYP smearing because its 
highly improved taste symmetry. 
In the $N_f=4$ flavor case I briefly studied the finite temperature phase transition to develop a feel for the parameters of the model. 
I calculated the  step scaling function at vanishing quark mass and  showed that the 2-lattice method
works equally well in the fermionic system.

The $N_f=16$ flavor model  brings in new challenges as it is governed by an infrared fixed point at finite gauge coupling. At this IRFP the 
scaling dimension of the gauge coupling is small, it is an almost marginal operator. Accordingly the  gauge coupling runs very slowly. The 
measured step scaling function is consistent with an almost marginal operator, but on its own it cannot predict if the gauge coupling is relevant or irrelevant.
I argued that the evolution of blocked operators signal if the RG flow is towards an IRFP or to the $\b=0$ trivial FP. For the $N_f=16$ model the flow clearly indicates an IRFP.

Finally I presented preliminary measurements for the scaling dimension of the mass. I found $y_m=1.02(7)$,   undistinguishable from the 
free field exponent. This is not surprising for the $N_f=16$ theory.

\section{Acknowledgment }

I thank T. DeGrand and J. Kuti for discussions and T. DeGrand for his help with modifying the staggered nHYP code for many fermions. 
This research was partially supported by the US Dept. of Energy. 

\bibliographystyle{apsrev}
\bibliography{lattice}

\end{document}